# Usinage de poches en UGV – Aide au choix de stratégies

Kwamivi Mawussi\*,\*\* - Sylvain Lavernhe\* - Claire Lartigue\*,\*\*\*

\* *Laboratoire Universitaire de Recherche en Production Automatisée*
*ENS de Cachan – Université Paris Sud (Paris XI)*
*61 avenue du Président Wilson, F-94235 Cachan cedex*

*{mawussi, lavernhe, lartigue}@lurpa.ens-cachan.fr*

\*\* *IUT de Saint Denis – Université Paris Nord (Paris XIII)*
*place du 8 Mai 1945, F-93206 Saint Denis cedex*

\*\*\* *IUT de Cachan – Université Paris XI*
*9 avenue de la division Leclerc, F-94234 Cachan cedex*

RÉSUMÉ. *Les travaux proposés dans cette communication s'inscrivent dans une démarche générale de génération semi-automatique d'usinage de poches dans le domaine aéronautique. Nous nous intéressons plus spécifiquement à l'aide au choix de stratégie d'usinage pour une géométrie de poche dans un contexte d'Usinage à Grande Vitesse (UGV). Après avoir défini des critères géométriques permettant la classification des entités poches et des entités spécifiques « aéronautiques », nous nous intéressons à l'association aux entités d'un ensemble d'outils capables. L'outil capable génère une géométrie de zone à usiner. La dernière partie est consacrée à l'association d'une stratégie d'usinage optimale à une géométrie donnée. Les résultats montrent que les critères géométriques d'analyse doivent être complétés par une étude dynamique. Dans un contexte UGV, il convient d'intégrer les spécificités Machine/CN dans le calcul de trajectoire.*

ABSTRACT. *The paper deals with associating the optimal machining strategy to a given pocket geometry, within the context of High-Speed Machining (HSM) of aeronautical pockets. First we define different classes of pocket features according to geometrical criteria. Following, we propose a method allowing to associate a set of capable tools to the features. Each capable tool defines a machined zone with a specific geometry. The last part of the paper is thus dedicated to associate the optimal machining strategy to a given geometry within the context of HSM. Results highlight that analyses must be conducted in a dynamical as well as a geometrical viewpoint. In particular, it becomes necessary to integrate dynamical specifities associated to the behavior of the couple machine/NC unit in the tool path calculation.*

MOTS-CLÉS : *usinage de poches ; UGV ; entité poche ; stratégie d'usinage.*

**1. Introduction**

Dans le domaine aéronautique, une part importante des pièces de structure est constituée de cavités ou poches qu'il faut évider. L'usinage de ces poches nécessite en général l'enlèvement d'un important volume de matière en ébauche ou en ébauche et finition combinées. D'un point de vue géométrique, une poche est constituée d'un fond et d'une surface latérale (Hatna *et al.*, 1998) s'appuyant sur un contour ouvert ou fermé. Le contour et par suite la surface latérale comprend un contour externe et aucun ou plusieurs contours internes (Tang *et al.*, 2003). Les portions de surface latérale s'appuyant sur des contours internes sont appelées îlots.

Si l'on s'intéresse à l'optimisation du processus d'usinage en Usinage Grande Vitesse (UGV), deux problèmes majeurs apparaissent : l'un lié à l'usinabilité de l'ensemble des poches constituant la pièce, l'autre concernant l'association d'une stratégie d'usinage optimale vis à vis du critère temps d'usinage de la pièce. Replacé dans un contexte de génération automatique ou semi-automatique d'usinage, ces problèmes peuvent être abordés au travers de la définition d'une base de connaissance pour l'usinage de pièces aéronautiques.

Cette base de connaissance s'appuie sur la décomposition de la pièce en entités. Un processus de reconnaissance des entités peut ainsi être mis en place (Mawussi, 1995). A ce stade, seule la géométrie des entités est détectée, donc connue. La base de connaissance permet d'associer à chaque géométrie type une stratégie d'usinage, c'est-à-dire le mode d'usinage, les entrées/sorties matières, les outils ainsi que les paramètres d'usinage qui sont préconisés. Si le processus est entièrement automatique, la stratégie appliquée est celle qui aura été déclarée optimale dans la base de connaissance. Dans le cas d'une démarche semi-automatique, l'opérateur peut intervenir en modifiant au choix, la gamme, le mode d'usinage, etc… s'il souhaite appliquer certaines règles spécifiques ou " métier ". En appliquant ce processus à toutes les entités, la dernière étape consiste à relier les usinages des poches pour terminer l'usinage de la pièce. L'ordonnancement global est basé sur les entrées/sorties matière de chaque entité et minimise les trajets hors-matière, donc le temps d'usinage.

Les travaux que nous présentons ici s'inscrivent dans une démarche de génération semi-automatique d'usinage des pièces de type aéronautique, et concernent plus spécifiquement l'aide au choix de stratégie d'usinage pour une géométrie de poches dans un contexte UGV. Après avoir défini les classes d'entités sur lesquelles nous appuyons nos travaux, nous proposons une méthode d'association « d'outils capables » à la poche, selon un critère de productivité. L'analyse du lien entre stratégie d'usinage et géométrie type de poche est ensuite détaillée, intégrant les spécificités UGV, dans l'optique de définir une base de connaissance.

## 2. Analyse des poches

### 2.1. *Caractérisation des poches*

En usinage, une poche est une cavité présente sur une pièce. Par cavité, on désigne toute partie creuse de la pièce considérée comme un volume. D'un point de vue géométrique, les cavités considérées dans le domaine aéronautique sont composées d'un fond, d'une frontière latérale et éventuellement d'îlots (Figure 1a). La frontière latérale s'appuie sur un contour appartenant soit au fond de la poche, soit à la surface supérieure généralement plane (plan supérieur). L'ensemble de ces éléments géométriques permet d'évaluer le volume de matière à usiner.

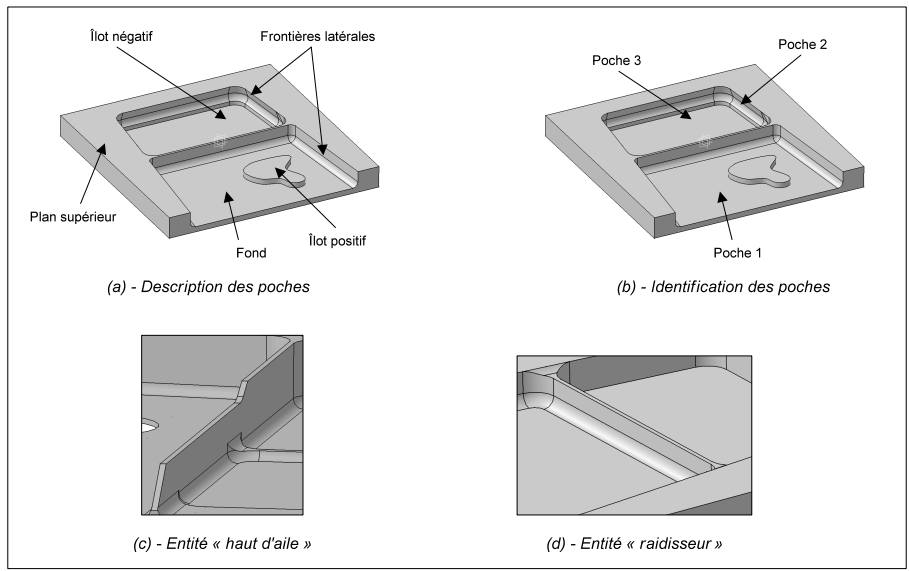

(a) - Description des poches    (b) - Identification des poches

(c) - Entité « haut d'aile »    (d) - Entité « raidisseur »

**Figure 1.** *Définition des poches dans le domaine aéronautique*

Les îlots sont définis à partir d'un contour fermé appartenant au fond de la poche. D'un point de vue topologique, toute section contenant un contour d'un îlot contient également un contour de la frontière latérale. Lorsque ce n'est pas le cas, l'îlot est dit négatif. Dans le cadre de notre travail, un îlot négatif est considéré comme une poche à part entière (Figures 1a et 1b). Sur les pièces réalisées dans le domaine aéronautique, apparaissent des formes (ou entités) spécifiques comme le « haut d'aile » ou le « raidisseur » (Figures 1c et 1d). Ces formes assurent les liaisons entre plusieurs poches et leur usinage pose des problèmes particuliers que nous aborderons plus loin.

Les différentes formes prises par chaque élément géométrique et les différentes configurations topologiques entre ces éléments géométriques constituent autant de critères de classification des poches. Le fond d'une poche peut être constitué d'une ou plusieurs surfaces, permettant ainsi de distinguer les poches à fond plan et celles à fond complexe. La frontière latérale peut être fermée ou non. Dans ce dernier cas, on parle de poche de coin ou de poche ouverte. Par ailleurs, cette frontière latérale peut être perpendiculaire ou non au fond de la poche. Cette configuration topologique contraint fortement le choix du type d'usinage à adopter. Plusieurs classifications ont été proposées dans les travaux antérieurs. Parmi ces travaux, on peut citer ceux de (Hatna *et al.*, 1998) et (Tseng *et al.*, 1994).

Les classifications proposées dans ces travaux ne sont pas adaptées au cadre de notre travail car elles n'intègrent pas le point de vue fabrication de la pièce. Les informations concernant l'usinage de la poche, le mode d'usinage associée à la géométrie, les techniques d'entrée et de sortie matière…, ne sont pas prises en compte. De plus le cadre d'Usinage à Grande Vitesse (UGV) impose des contraintes spécifiques (Pateloup *et al.*, 2003). Il est donc nécessaire d'introduire des critères d'ordre technique pour optimiser l'évidemment des poches. Les classes d'entités retenues dans le cadre de notre travail sont données dans la Figure 2. Les critères d'ordre technique viennent en complément des critères géométriques associés aux classes d'entités poche.

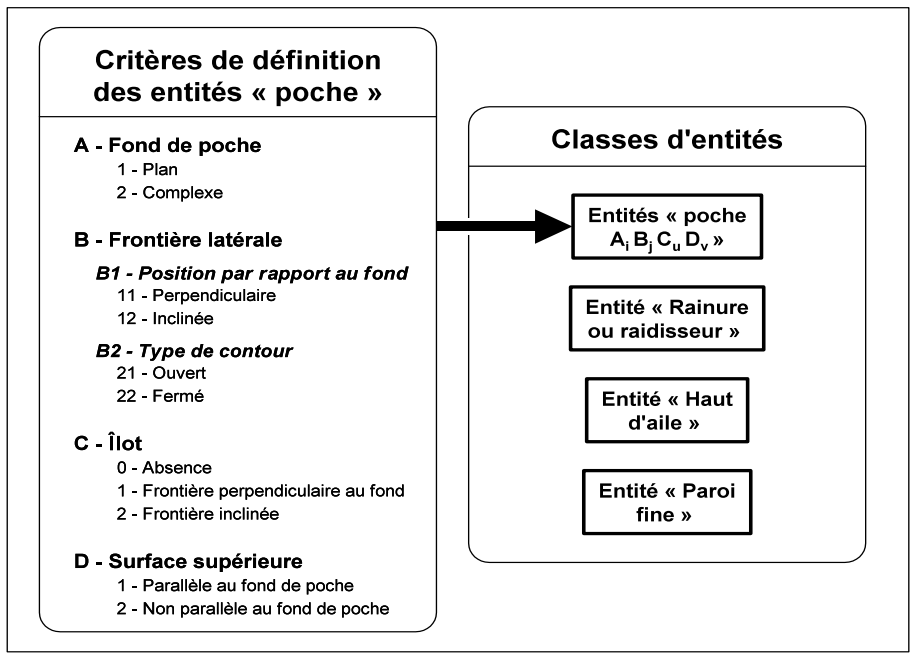

**Figure 2.** *Classification des entités*

**2.2.** *Décomposition des poches*

Une fois une poche identifiée, elle doit être décomposée en zones d'usinage. Cette décomposition se fait suivant 2 critères : diamètre d'outil et entité spécifique. Les entités spécifiques comme la « paroi fine » nécessitent des gammes et stratégies d'usinage particulières en général pour éviter leur déformation ou rupture. En effet, il faut usiner les zones avoisinantes de ces entités spécifiques progressivement de part et d'autre de la forme finie. A ce niveau, la décomposition des poches conduit à définir une zone non usinée autour des entités spécifiques. Le reste de la poche ainsi dissociée de la zone non usinée peut être traité comme toute entité de base. La définition des zones non usinées ne sera pas présentée dans cette communication.

La décomposition suivant le critère diamètre d'outil permet de séparer une poche en zones correspondant à des diamètres d'outils différents. L'ensemble des zones ainsi séparées dans une poche doit permettre de réduire considérablement le temps total d'usinage. Le choix des diamètres d'outils associés à ce type de décomposition a fait l'objet de plusieurs travaux parmi lesquels on peut citer ceux de (Yao *et al.*, 2003) et (Veeramani *et al.*, 1997). Dans le cadre de notre travail, nous proposons une méthode de détermination des diamètres d'outils par dichotomie et itérations.

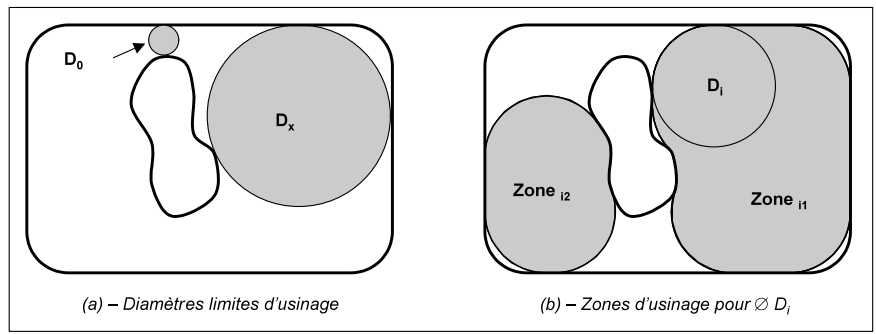

*(a) – Diamètres limites d'usinage*   *(b) – Zones d'usinage pour ⌀ $D_i$*

**Figure 3.** *Décomposition d'une poche en zones d'usinage*

Etant donnée une poche, on désigne par $D_0$ et $D_x$ respectivement les diamètres capable et maximal d'usinage (Figure 3a). Le diamètre capable est celui du plus grand outil permettant d'usiner la poche sans dégagement intermédiaire. Le diamètre maximal d'usinage $D_x$ est celui du plus grand outil pouvant être introduit dans la poche. Ces deux diamètres constituent les valeurs initiales de la méthode de décomposition de la poche considérée. Une méthode de détermination des diamètres $D_0$ et $D_x$ a été proposée par (Mawussi *et al.*, 2000).

Tout diamètre $D_i$ ayant une valeur intermédiaire à celles de $D_0$ et $D_x$ divise la poche en un ensemble de zones usinées $Z_{ij}$ et un ensemble de zones non usinées (Figure 3b). Pour chaque diamètre $D_i$, nous calculons le temps d'enlèvement de matière $T_i$, de la façon suivante : $T_i = L_i/V_c$, où $L_i$ est la longueur du trajet et $V_c$, la vitesse capable définie par la base de données outil.

Le principe de la méthode de décomposition est le suivant. A l'issue de la première itération, on dispose de trois diamètres : $D_0$, $D_x$, et $D_{0x}$. Ce dernier est obtenue par dichotomie ($D_{0x} = \frac{1}{2}[D_0 + D_x]$) et l'ensemble des trois diamètres donne deux intervalles de variation ([ $D_0$ , $D_{0x}$ ] et ] $D_{0x}$ , $D_x$ ]). A ce niveau, la comparaison du débit de matière enlevée (rapport entre le volume de matière enlevée et le temps nécessaire) de chacun des outils (Figure 4) permet d'éliminer ou non les intervalles de variation (un seul intervalle peut être éliminé à la première itération).

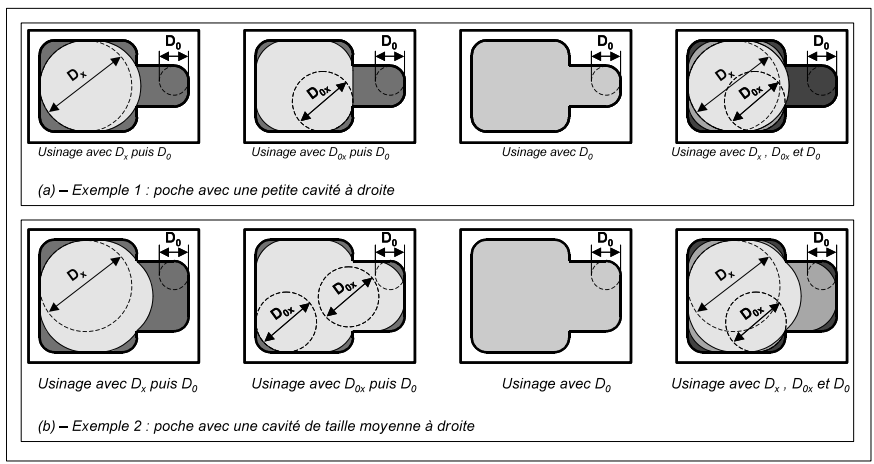

**Figure 4.** *Principe de décomposition*

La Figure 4 montre deux exemples de comparaison de volumes de matière enlevée. L'accessibilité de la cavité de droite par l'outil de diamètre $D_{0x}$ (Figure 4b) modifie de façon importante les temps d'enlèvement de matière et par suite le débit de matière enlevée. Le choix des intervalles de variation du diamètre d'outil à retenir est fait à partir d'un seuil de débit de matière égal à 5%. Ainsi, dans le cas de l'exemple 1 de la figure 4, le débit associé au diamètre $D_{0x}$ n'est pas supérieur de 5% à celui associé au diamètre $D_0$. L'intervalle ] $D_{0x}$ , $D_x$ ] est donc éliminé. Le cas de l'exemple 2 de la figure 4 est différent. Le débit associé au diamètre $D_{0x}$ est supérieur de près de 30% à celui associé au diamètre $D_0$. Les deux intervalles [ $D_0$ , $D_{0x}$ ] et ] $D_{0x}$ , $D_x$ ] sont retenus pour la deuxième itération.

La valeur du seuil de débit de matière est fixée à partir des règles métier. Elle peut être affinée à l'aide d'une formalisation analytique de l'expression du débit de matière. Par ailleurs, l'intégration d'autres paramètres, comme ceux liés à l'UGV et à la qualité géométrique des formes usinées, dans la méthode est actuellement en cours de validation. La méthode ainsi étendue fera l'objet d'une communication ultérieure.

A l'issue de cette étape, nous pouvons associer un ensemble d'outils capables à l'entité poche considérée. Nous nous intéressons dans la suite au lien entre géométrie de poche et stratégie d'usinage optimale pour un outil donné.

**3. Analyse des stratégies d'usinage de poches en UGV**

Il existe plusieurs stratégies d'usinage qui ont fait l'objet de travaux antérieurs et qui sont intégrées dans la plupart des outils de Fabrication Assistée par Ordinateur (FAO). Une stratégie d'usinage définit le mode d'usinage et les entrée-sortie de l'outil. Dans un contexte UGV, la génération des trajets d'outils pour les poches s'accompagne de la définition de paramètres supplémentaires permettant de garantir la continuité lors du suivi de trajectoire.

Afin d'analyser l'influence des différents paramètres caractérisant la stratégie d'usinage sur le temps d'usinage, pour une géométrie donnée, nous avons défini une géométrie type présentée Figure 5. Cette géométrie est significative des formes de poches rencontrées dans le domaine aéronautique. La stratégie d'usinage optimale est donc celle qui conduit au temps d'usinage le plus faible.

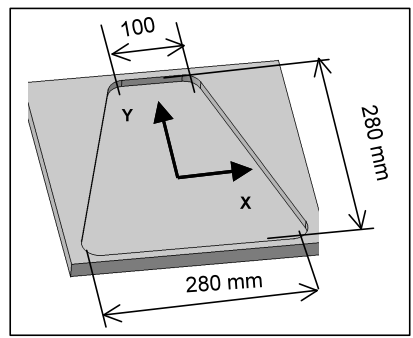

**Figure 5.** *Géométrie de la « poche test »*

**3.1.** *Stratégie d'usinage*

Comme nous l'avons vu précédemment, pour une géométrie de poche donnée et un outil capable associé, une stratégie d'usinage est définie par la donnée :

- du mode de parcours ou de balayage de la poche,
- des trajets d'entrée-sortie matière de l'outil,
- des paramètres opératoires,
- des paramètres spécifiques UGV.

Les paramètres opératoires (vitesse d'avance, tolérance d'usinage, prise de passe….) sont en général définis par la base de données COM ainsi que par les spécifications géométriques de défaut de forme et d'état de surface. Nous les supposerons donnés et constants pour notre étude.

Les modes de parcours les plus courants sont l'usinage hélicoïdal (ou spirale) et l'usinage parallèle à une direction. Dans ce dernier cas, le balayage est le plus souvent réalisé en mode aller-retour (figure 6a).

Les trajets entrée-sortie matière peuvent être de différentes natures : selon l'axe de l'outil, approche tangentielle à la surface , en rampe, … . Cependant, on ne retiendra ici que les trajets les plus couramment utilisés dans le domaine aéronautique (Figure 6b) : entrée tangentielle par le flanc pour les poches ouvertes, pénétration en spirale pour les poches fermées. Les trajets de sortie sont définis par le graphe de connexion des poches (Lavernhe, 2003).

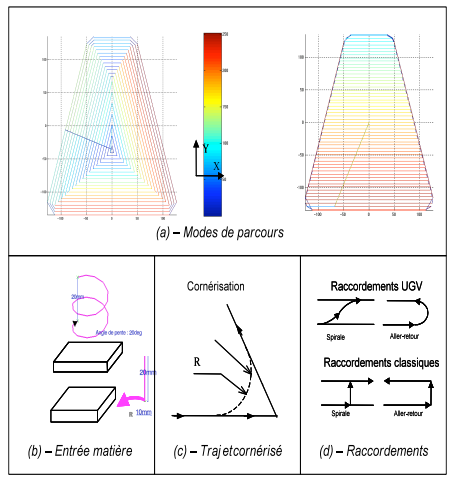

**Figure 6.** *Caractéristiques associées à une stratégie d'usinage*

En ce qui concerne les paramètres spécifiques UGV, ils permettent de garantir un mouvement continu et fluide (sans à-coup) de l'outil en cours d'usinage. En effet, la trajectoire, lorsqu'elle est décrite sous un format linéaire, est constituée d'un ensemble de segments de droite raccordés $C^0$ entre eux. Chaque changement de direction important entraîne une chute de vitesse, ce qui conduit à une diminution du temps d'usinage mais aussi dans certains cas à des marques sur la pièce. Pour pallier à ce problème, les directeurs de Commande Numérique (CN) ont intégré des fonctionnalités qui lissent les trajectoires, soit par ajout d'arcs de cercle au franchissement des discontinuités, soit en approximant la trajectoire par une fonction polynomiale (Figure 7) (Dugas, 2002).

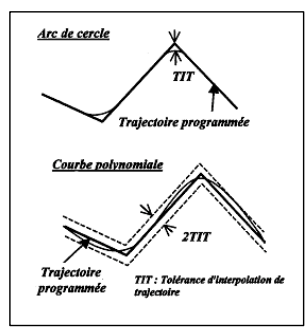

**Figure 7.** *Transformation de la trajectoire en temps réel (Dugas, 2003)*

Cette fonctionnalité est désormais intégrée dans certains logiciels de FAO qui proposent des fonctions dites de « cornérisation » des angles aigus, c'est à dire l'ajout d'un arc de cercle, de rayon paramétrable au niveau des discontinuités (Figure 6c). De même, les changements de passe classiques imposent des variations brusques de direction et peuvent être remplacés par des raccordements de type UGV, dont on contrôle les dimensions et la courbure (figure 6d). L'ensemble des paramètres détaillés ci-dessus définissent les paramètres spécifiques UGV. Nous nous intéressons dans la partie suivante à l'influence de certains des paramètres sur le temps d'usinage.

**3.2.** *Analyse de la variation des caractéristiques d'usinage*

Pour les essais que nous avons retenus, les trajets entrée-sortie matière sont choisis selon les critères définis au paragraphe précédent. Les paramètres opérateurs retenus sont les suivants : $V_f$ = 10 m/min ; l'outil est une fraise 2T $\phi$ 10 mm ; la prise de passe radiale est de 5 mm.

Les facteurs d'influence que nous avons choisi d'étudier sont le choix du mode de parcours, spirale ou aller-retour, avec dans chacun des cas les raccordements classiques et les raccordements UGV. Pour chacun des 4 essais, nous analysons le temps d'usinage. Cependant, le temps d'usinage FAO, identique dans chacun des cas, n'est pas significatif du temps réel d'usinage. En effet, le temps d'usinage calculé suppose que l'ensemble du trajet est parcouru à la vitesse programmée. Or, la vitesse programmée n'est atteinte que sur les portions de trajet de grande longueur. Ainsi, le temps d'usinage est en partie conditionné par le rapport entre petits et grands segments élémentaires.

Pour réaliser notre analyse, il convient donc de réaliser la simulation de l'usinage sur machine, afin de mesurer les temps d'usinage effectifs. Nos essais ont été réalisés sur une machine Mikron UCP 710 équipée d'une CN Siemens 840D. Les temps mesurés pour chacun des essais sont reportés sur le tableau 1.

| Mode d'usinage | Spirale | Aller-retour | Spirale UGV | Aller-retour UGV |
|---|---|---|---|---|
| **Temps mesuré** | 1 min 43.4 s | 1 min 43.3 s | 1 min 55.6 s | 1 min 50.9 s |
| **Temps estimé en FAO** | 30 s | 32 s | 30 s | 32 s |

**Tableau 1.** *Temps d'usinage mesurés*

Les résultats mettent en avant que les temps associés aux trajectoires UGV sont plus importants que les temps associés aux trajectoires classiques. Ce résultat est en concordance avec l'analyse de la longueur des trajets élémentaires présentés sur la figure 8. En effet, afin d'effectuer les essais en interpolation linéaire, les arcs de cercles sont discrétisés, ce qui entraîne une augmentation importante des segments de petite longueur.

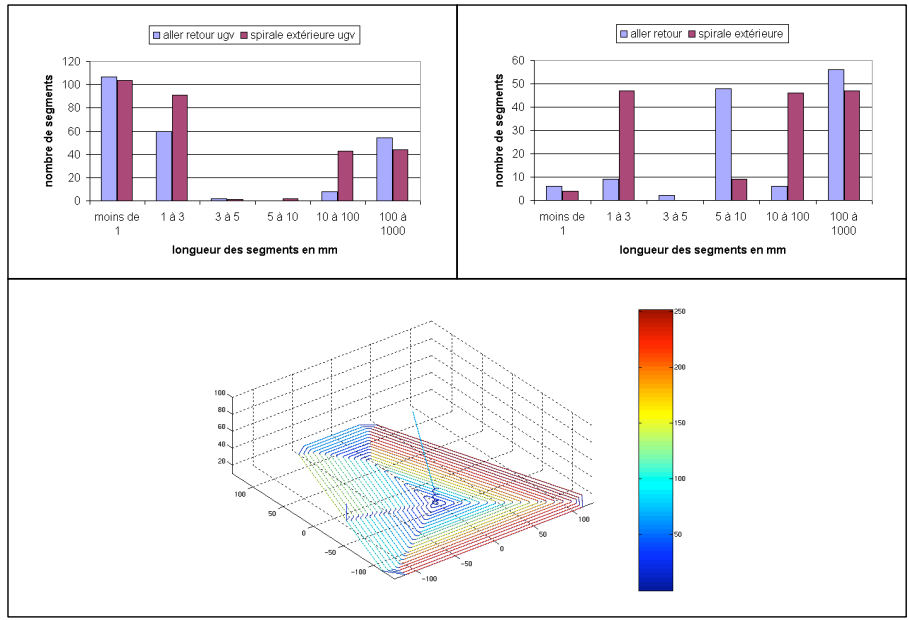

**Figure 8.** *Résultats*

De même, la cornérisation n'est ici pas concluante étant donné que le rayon des arcs de cercle n'a pas été optimisé en fonction des capacités du couple machine/directeur de commande numérique. La vitesse programmée ne peut être atteinte sur un cercle que si le rayon est tel que $r > (V_f)^2/a_{maxi}$, où $a_{maxi}$ est l'accélération maximale associée aux axes de la MOCN.

Ainsi, il est important de noter que les trajectoires ne peuvent plus être générées en FAO, sans avoir au préalable une bonne connaissance du couple machine/CN sur lequel l'usinage doit être réalisé. Cependant, les évolutions sont telles qu'il est difficile d'appréhender de manière simple le comportement de la CN en suivi de trajectoire.

Si l'on connaît les lois de commande, mettant en avant les paramètres $a_{maxi}$, jerk, etc., …, d'autres paramètres sont programmables et ont une influence considérable sur le comportement de la CN (anticipation en vitesse (figure 9c), modes d'accélération brisk ou soft (figure 9b), ….).

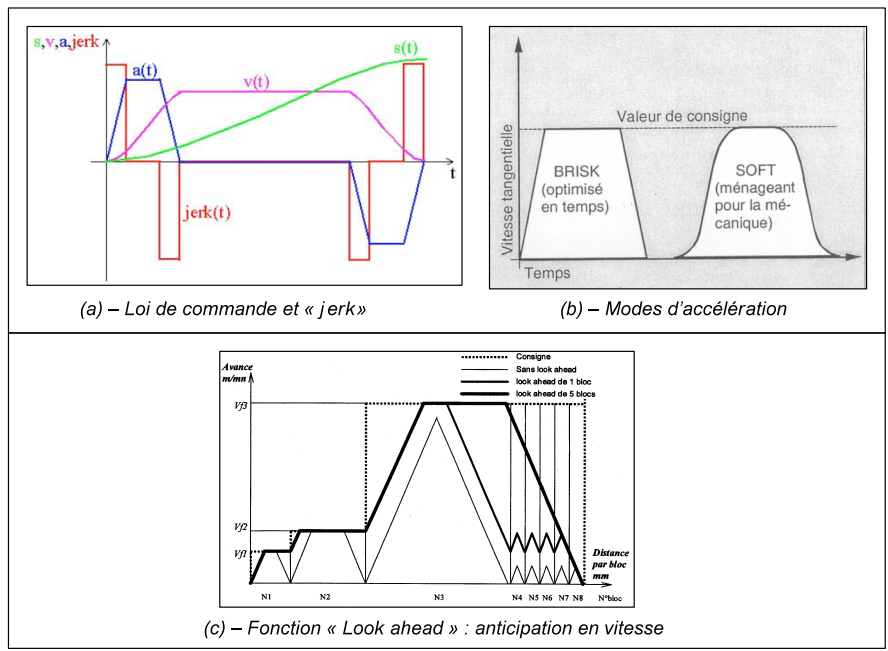

(a) – Loi de commande et « jerk »    (b) – Modes d'accélération

(c) – Fonction « Look ahead » : anticipation en vitesse

**Figure 9.** *Paramètres liés à la commande numérique*

L'étude de la géométrie seule des trajectoires n'est pas suffisante pour définir des critères significatifs pouvant conduire au choix de stratégie optimale. L'étude doit être complétée par une étude dynamique plus approfondie.

On retiendra cependant que la stratégie doit être définie par les entrées-sorties matière, le mode de parcours et les raccordements entre passes. Les entrées-sorties sont fonctions de la géométrie de la poche (ouverte ou fermée) et des connections entre poches. Le mode de parcours sera choisi de manière à minimiser la proportion entre le nombre de segments de faible longueur et le nombre de segments de grande longueur. Enfin, les raccordements sont choisis de manière à maximiser les rayons en fonction de l'accélération maximale de la machine.

**4. Conclusion**

Nous avons présenté dans cette communication une approche d'aide au choix de stratégies d'usinage de poches. Les travaux réalisés s'inscrivent dans un contexte de fabrication de pièces spécifiques au domaine de l'aéronautique. Dans un premier temps, nous avons analysé les caractéristiques essentiellement géométriques des poches. Cette analyse a d'abord conduit à une classification des entités pour ensuite déboucher sur la proposition d'une méthode de décomposition des entités de type poche. Cette méthode devrait être complétée par l'introduction de paramètres spécifiques à l'UGV. L'intégration de ces paramètres ainsi que la validation finale de la méthode sont actuellement en cours.

Nous avons présenté dans la seconde partie de la communication une analyse des différentes caractéristiques des stratégies d'usinage. Ces caractéristiques couvrent les modes de parcours du volume de matière enlevée dans les poches, les trajets d'entrée-sortie matière, ainsi que d'autres paramètres liés à l'UGV comme les raccordements de trajets, les transformations de trajets et la modification des points anguleux. Les résultats de cette analyse constituent la base de connaissance à partir de laquelle est opérée l'aide au choix des stratégies d'usinage. Les travaux futurs s'orientent vers la création d'un outil d'aide au choix des stratégies d'usinage.

**5. Bibliographie**

Dugas A., CFAO et UGV – Simulation d'usinage de formes complexes, Thèse de Doctorat, Université de Nantes - École Centrale de Nantes, France, 2002.

Hatna A., Grieve R. J., Broomhead P., « Automatic CNC milling of pockets : geometric and technological issues », *Computer Integrated Manufacturing Systems*, vol. 11, n° 4, 1998, p. 309-330.

Lavernhe S., Usinage semi-automatique des pièces aéronautiques de structure : Aide au choix de stratégies d'usinage, Mémoire de Recherche, DEA de Production Automatisée, ENS de Cachan, France, 2003.

Mawussi K., Modèle de représentation et de définition d'outillages de formes complexes. Application à la génération automatique de processus d'usinage, Thèse de Doctorat, ENS de Cachan, France, 1995.

Mawussi K., Duong V.-H., Kassegne K., « Determination of the parameters of cutting tools in integrated design of products », *7$^{th}$ ISPE International Conference on Concurrent Engineering: Research and Applications CE2000*, Lyon, July 17-20, 2000, p. 465-472.

Pateloup V., Duc E., Lartigue C., Ray P., « Pocketing optimization for HSM, geometry tool path and interpolation mode influence on dynamic machine tool behavior », *Machine Engineering*, vol. 3 (1-2), 2003, p. 127-138.

Tang K., Joneja A., « Traversing the machining graph of a pocket », *Computer-Aided Design*, vol. 35, 2003, p. 1023-1040.

Tseng Y. J, Joshi S. B., « Recognizing multiple interpretations in 2 ½ D machining of pockets », *International Journal of Production Research*, vol. 32, n°5, 1994, p. 1063-1086.

Veeramani D., Gau Y.-S., « Selection of an optimal set of cutting-tool sizes for 2½D pocket machining », *Computer-Aided Design*, vol. 29, n°12, 1997, p. 869-877.

Yao Z., Gupta S. K., Nau D. S., « Algorithms for selecting cutters in multi-part milling problems », *Computer-Aided Design*, vol. 35, 2003, p. 825-839.